\documentclass[11pt]{article}
\usepackage[T1]{fontenc} 
\usepackage{authblk}
\usepackage{geometry}
\usepackage{amsmath}
\usepackage{amssymb}
\usepackage{graphicx}
\usepackage{braket}
\usepackage{color}
\usepackage{mathrsfs}
\usepackage{hyperref}
\usepackage{setspace}
\usepackage{tensor}
\usepackage{isomath}
\usepackage{pgf,tikz}
\usepackage{mathrsfs}
\usetikzlibrary{arrows}
\definecolor{ffqqqq}{rgb}{1.,0.,0.}
\definecolor{xdxdff}{rgb}{0.49019607843137253,0.49019607843137253,1.}
\definecolor{ffqqff}{rgb}{1.,0.,1.}
\newcommand{\minus}{\backslash}
\DeclareMathOperator{\Ar}{Area}

\title{Holographic Inequalities and Entanglement of Purification}
\author{Ning Bao and Illan F. Halpern} 
\affil{Berkeley Center for Theoretical Physics, Berkeley, California}
\date{\today}

\begin{document}

\maketitle

\begin{abstract}
    We study the conjectured holographic duality between entanglement of purification and the entanglement wedge cross-section. We generalize both quantities and prove several information theoretic inequalities involving them. These include upper bounds on conditional mutual information and tripartite information, as well as a lower bound for tripartite information. These inequalities are proven both holographically and for general quantum states. In addition, we use the cyclic entropy inequalities to derive a new holographic inequality for the entanglement wedge cross-section, and provide numerical evidence that the corresponding inequality for the entanglement of purification may be true in general. Finally, we use intuition from bit threads to extend the conjecture to holographic duals of suboptimal purifications.
\end{abstract}

\section{Introduction}
There has been much recent interest in the interplay between the fields of quantum information and quantum gravity. One central point of interest is on the discussions of notions of entanglement measures in the context of the AdS/CFT correspondence \cite{malda, Witten:1998qj}. In particular, the holographic formula relating entanglement entropy to bulk area of a boundary homologous minimal surface by Ryu and Takayanagi \cite{Ryu:2006ef,Ryu:2006bv} (later extended to extremal surfaces in the covariant case by \cite{Hubeny:2007xt}) has spurred a great deal of interest, from its ability to constrain what sets of states can be dual to classical bulk gravity theories \cite{EntCone, Hayden:2011ag} to its role as motivation for the idea that the gravity theory is emergent from the entanglement properties of the boundary field theory \cite{VanRaamsdonk:2010pw,Maldacena:2013xja, Faulkner:2013ica, Swingle:2014uza}.

Even more recently work has been done \cite{Nguyen,TakUme} conjecturing that a holographic object, the entanglement wedge cross section $E_W$ separating two regions, is dual to the information theoretic concept of the entanglement of purification $E_p$. This conjecture, which we refer to as the $E_W=E_p$ conjecture, was made on the basis that $E_W$, a holographic object, obeys the same set of inequalities that $E_p$ is known to obey. This would be a compelling correspondence, as it is not known how to calculate $E_p$ for generic quantum states, whereas $E_W$ is an often finite geometric quantity that is simply calculable.

In this work, we will study and generalize the relationship between $E_W$, $E_p$, and the holographic entanglement entropy inequalities in three ways: first, we investigate whether $E_W$ can nontrivially bound combinations of entanglement entropies that appear in the holographic entropy inequalities; second, we check whether $E_p$ provably provides the same type of bounds to these objects; lastly, we ask whether one can extend the $E_W=E_p$ conjecture to suboptimal purifications and cuts of the entanglement wedge. We will find that the answers to all three of these questions appear to be affirmative, thus providing more evidence for the $E_W=E_p$ conjecture of \cite{Nguyen,TakUme}.

\section{Review of known results}
\subsection{Basic properties of $E_W$ and $E_p$}
\begin{figure}[h]
    \centering
    \includegraphics[trim={3cm 5cm 9cm 4cm},clip,width=0.85\textwidth]{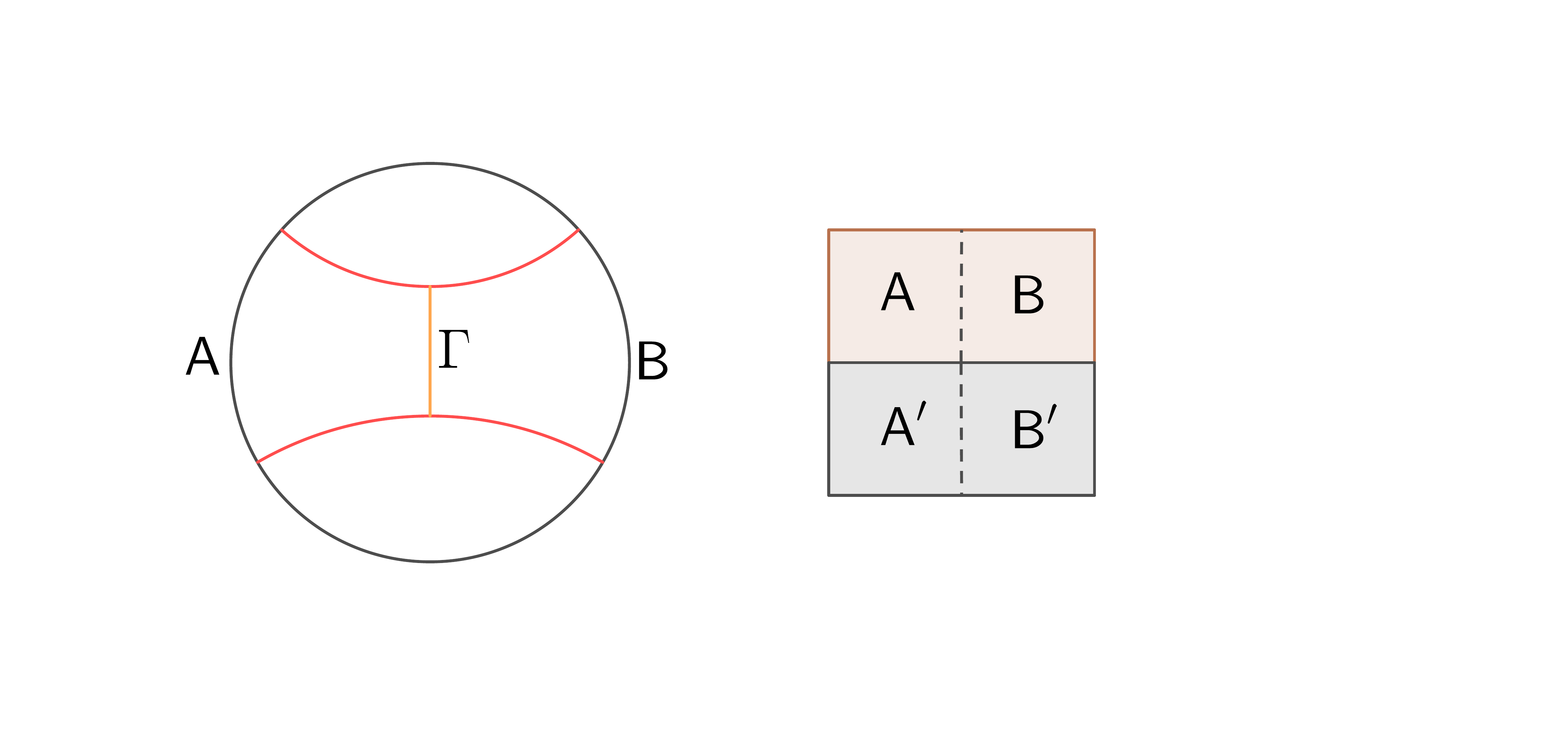}
    \caption{To the left, $\Gamma$ is the minimal surface the separates the entanglement wedge cross-section of $AB.$ Its area is $E_W[A:B].$ To the right, $A'$ and $B'$ purify $AB.$ For a choice of $A'$ and $B'$ over all such purifying systems that minimizes the entanglement across the dashed partition we have $E_p(A:B)=S(AA').$}
    \label{fig:EwEp}
\end{figure}

Let us define both the entanglement wedge cross section $E_W,$ and the entanglement of purification $E_p.$. First, we define holographic states to be quantum states of the boundary conformal field theory that are dual to a well defined classical bulk gravitational theory in AdS/CFT.  For a holographic state, the {\bf entanglement wedge cross-section} is defined for any two regions of time reversal symmetric slices (though the generalization to the fully covariant case exists in \cite{TakUme}) as
\begin{equation}
E_{W}(A:B) =
\min \{\Ar(\Gamma); \Gamma \subset r_{AB} \text{ splits } A \text{ and } B\}
\end{equation}
where $r_{AB}$ is the entanglement wedge \cite{Headrick:2014cta} of $AB=A \cup B$ (see Figure \ref{fig:EwEp}). In words, $E_W$ is the minimal area of a surface $\Gamma$ that splits $r_{AB}$ into two regions, one of which is bounded by $A$ but not $B,$ and other by $B$ but not $A.$ If $s_A, s_B$ and $s_{AB}$ denote Ryu-Takayanagi (RT) surfaces, then we want $\Gamma$ to split $r_{AB}=r_{AB}^{(A)} \sqcup r_{AB}^{(B)}$ (here $\sqcup$ denotes disjoint union) and $s_{AB}=s_{AB}^{(A)} \sqcup s_{AB}^{(B)}$ with $\partial r_{AB}^{(A)}=A \cup s_{AB}^{(A)} \cup \Gamma.$\footnote{It is also worth noting that $E_W$ (and its to-be-developed generalization) is finite if the regions being split are nonadjacent in the boundary theory.} In this work, we refer interchangeably to the surface and the area thereof as the entanglement wedge cross-section, but the meaning should be clear from context.

Now, consider an arbitrary bipartite quantum system $AB.$ The {\bf entanglement of purification} $E_p(A:B)$ is defined by 

\begin{equation}
E_p(A:B)=
\min \{S(AA');AA\rq{}BB\rq{} \text{pure} \}
\end{equation}
where $S$ is the Von Neumann entropy (see Figure \ref{fig:EwEp}). Note that because the overall state is pure this is symmetric under $A\leftrightarrow B$.

Both $E_W$ and $E_p$ are known to satisfy the following inequalities:
\begin{eqnarray}
\min (S_A, S_B)\geq E(A:B) \geq \frac{1}{2} I(A:B) \label{eq:Eulb}\\
E(A:BC) \geq E(A:B) \label{eq:Esub} \\
E(AB:C) \geq \frac{1}{2} \left(I(A:C) + I(B:C)\right) \label{eq:Emon},
\end{eqnarray}
where $E$ here can stand for either $E_p$ or $E_W,$ and $I(A:B)\equiv S(A)+S(B)-S(AB)$ is the mutual information between $A$ and $B$. We refer the reader to \cite{TakUme} for clear proofs of these inequalities in the context of $E_W$, and \cite{Terhal,Bagchi} for the same for $E_p$. These coinciding bounds for $E_p$ and $E_W$ is what motivated the conjecture of \cite{Nguyen, TakUme} that $E_W$ is the holographic dual of $E_p.$

\subsection{Entanglement Entropy Inequalities}

Before we study potential new inequalities for $E_p$ and $E_W$, let's list some known inequalities for entanglement entropy that will prove useful in the upcoming discussion. For holographic proofs of these inequalities we refer the reader to \cite{EntCone, Hayden:2011ag, HeadTak}. All tripartite quantum states satisfy strong subadditivity (SSA):

\begin{equation}
I(A:B|C) \equiv S(BC)+S(AC)-S(ABC)-S(C) \geq 0,
\end{equation}
where $I(A:B|C)$ is the conditional mutual information. When $C=\emptyset,$ this reduces to subadditivity, or positivity of the mutual information. Moreover, all \emph{holographic} states satisfy monogamy of mutual information (MMI) \cite{Hayden:2011ag}:

\begin{equation}
I(A:BC) \geq I(A:B) + I(A:C) \Leftrightarrow I(A:B:C) \equiv I(A:BC)-  I(A:B) - I(A:C)\geq 0,
\end{equation}
where $I(A:B:C)$ is the tripartite information, which is symmetric under permutations of its arguments.

It is worth stressing that not all quantum states satisfy MMI. For example, the GHZ state defined by $\ket{\text{GHZ}}=\frac{1}{\sqrt{2}} \left(\ket{0}^{\otimes n} + \ket{1}^{\otimes n} \right)$ does not do so for $n\geq 4$.

Recently, several further holographic entanglement entropy inequalities were proven \cite{EntCone}. Among them, there's an infinite family of cyclic inequalities given by

\begin{equation}
C_{k}(A_1, \dots, A_{n}) \equiv \sum_{i=1}^n S(A_i | A_{i+1} \dots A_{i+k}) - S(A_1 \dots A_n) \geq 0 , \label{eq:cyc}
\end{equation}
where $n=2k+1,$ the indices are interpreted mod $n,$ $S(A|B)=S(AB)-S(B)$ is the conditional entropy, and $C_k$ is what we call the $k$-cyclic information (or just cyclic information). For $k=1,$ Eq.~(\ref{eq:cyc})  gives MMI, but for $k>1$ it gives a family of new and independent inequalities.

\section{Bounding holographic entanglement entropy with $E_W$}
\subsection{Generalized $E_W$}
In order to bound holographic entanglement entropy with $E_W$, we first slightly generalize the notion of entanglement wedge cross-section. The {\bf generalized entanglement wedge cross-section} $E^G_W$ is defined as
\begin{equation}
E^G_{W}(A:B) =
\min \{\Ar(\Gamma);\Gamma \subset r_{AB}-r_{A \cap B} \text{ splits } A \backslash B \text{ and } B\backslash A \}, 
\end{equation}
so now the surface $\Gamma$ separates $A\backslash B$ from $B\backslash A$ in the region defined by the entanglement wedge of $AB$ with the entanglement wedge of $A \cap B$ removed\footnote{For an illustration of what this generalization means geometrically, see Fig. \ref{fig:ssa}, in which $E^G_W(AC:BC)=\Ar(\Gamma)$)}. Note that if $A \cap B = \emptyset$ then $E^G_W(A:B)=E_W(A:B).$

We also define a convenient form of mutual information, $I^G(A:B)=I(A \backslash B,B \backslash A)=S(A \backslash B) + S(A \backslash B) - S(A\backslash B \cup B \backslash A).$ Similiarly, if the intersection between $A$ and $B$ is trivial this reduces to $I(A:B)$.
\subsection{$E^G_W$ obeys known $E_W$ inequalities}
This generalized entanglement wedge cross-section obeys the known inequalities for $E_W$ in Eqs.~(\ref{eq:Eulb}--\ref{eq:Emon}).

The upper bound in Eq.~(\ref{eq:Eulb}) follows from
\begin{eqnarray}
E^G_W(A:B) \leq E_W(A:B \backslash A) \leq \min (S(A),S(B \backslash A) )  \text{and} \nonumber\\ 
E^G_W(A:B) \leq E_W(A \backslash B :B) \leq \min (S(A \backslash B),S(B)),
\end{eqnarray}
where the first inequalities above follow from the fact that for $E^W_p(A:B),$ the minimum area curve $\Gamma$ that separates $A \backslash B$ from $B \backslash A$  in $r_{AB}-r_{A \cap B}$ so it can be no longer than optimal curve separating them in $r_{AB}.$

The lower bound follows from $r_{AB \backslash (A\cap B)} \subset (r_{AB}-r_{A\cap B}),$ which is a consequence of entanglement wedge nesting (EWE) \cite{Akers:2016ugt} and implies
\begin{eqnarray}
E_W^G(A:B) \geq E_W(A\backslash B:B \backslash A) \geq \frac{1}{2} I^G (A:B). \label{eq:EWGub}
\end{eqnarray}
It follows from entanglement wedge nesting that if $A \cap C = \emptyset,$ then
\begin{equation}
E_W^G(A:BC) \geq E_W^G (A:B).
\end{equation}
Finally, 
\begin{equation}
E_W^G(A:BC) \geq \frac{1}{2} \left( I^G(A:B) + I(A\backslash B:C)\right)
\end{equation}
follows from Eq.~(\ref{eq:EWGub}), MMI, and the disjointedness of $A$ and $C.$

\subsection{Upper bounding holographic conditional mutual information}
One can ask the question of whether or not holography in general, and $E^G_W$ in particular, provides an upper bound to the conditional mutual information. We note that this question was first answered in the affirmative by \cite{FreedHead} using bit threads, but it is instructive to treat it again here.

The holographic bound for $I(A:B|C)$ in \cite{FreedHead} reads:
\begin{equation}
I(A:B|C) \leq 2 E^G_W(AC:BC). \label{eq:flipssa}
\end{equation}
Note that in the case where $C=\emptyset,$ this reduces to $I(A:B) \leq 2 E_W(A:B).$

\begin{figure}[h]
\centering
\includegraphics[trim={0cm 9cm 10.5cm 5cm},clip,width=\textwidth]{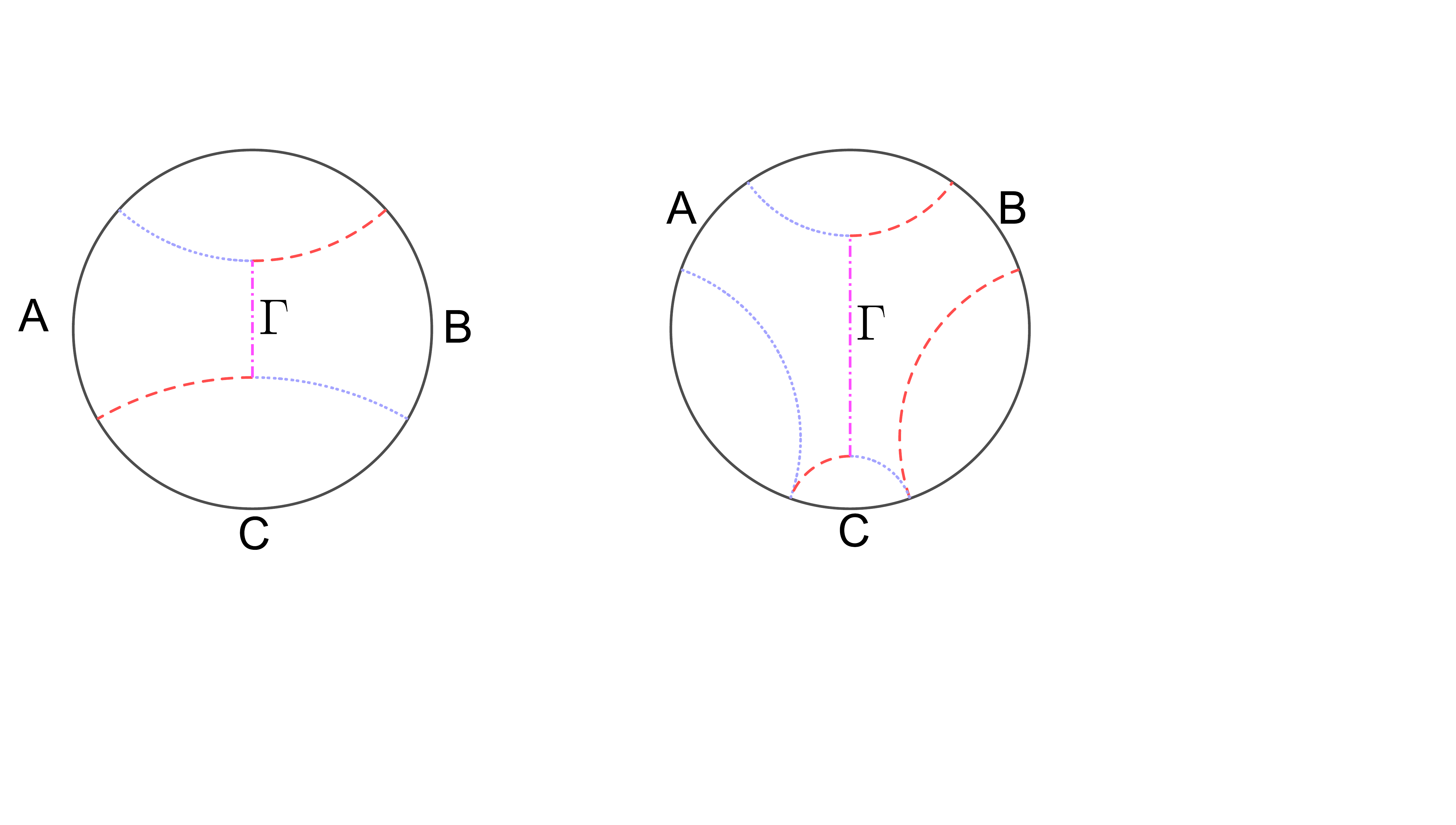}
\caption{Graphical proof of the upper bound on the conditional mutual information for both the case in which the regions A, B, and C are contiguous and the case in which they are disconnected. It is clear from the diagrams,and Ryu-Takayanagi, that the area of the dotted surfaces plus the area of the dash-dotted surface is greater than or equal $S(AC)$ and that the area of the dashed surfaces plus the area of the dash-dotted surface is greater than or equal $S(BC).$ Adding these two inequalities gives us the desired bound.}
\label{fig:ssa}
\end{figure}

This upper bound can also be proven using exclusion/inclusion \cite{HeadTak} or equivalently graph contraction \cite{EntCone}, with the main new technique used being that the cutting and regluing procedure is no longer constrained to only boundary anchored minimal surfaces, but potentially includes bulk-anchored minimal surfaces such as the entanglement wedge cross-section as well. See Figure \ref{fig:ssa}.

Let's now follow \cite{HeadTak} in putting into equations what Figure \ref{fig:ssa} shows. Let $s_X$ denote the RT surface of some boundary region $X,$ and $r_X$ denote its entanglement wedge so that the boundary of $r_X$ is $\partial r_X = X \cup s_X.$ 

Let $A, B,$ and $C$ be disjoint regions, and let $R=r_{ABC} \minus r_C.$ By EWE, $\partial R = s_{abc} \cup s_c.$ Let $\Gamma$ be the surface that satisfies the minimization in $E_W[AC:BC].$ Then, it splits $R$ into two disjoint regions $R^{(A)}$ and $R^{(B)}$ such that 

\begin{eqnarray}
\partial R^{(A)} &=& A + \Gamma + s_{ABC}^{(A)}+s_C^{(A)} \nonumber \\
\partial R^{(B)} &=& B + \Gamma + s_{ABC}^{(B)}+s_C^{(B)}, 
\end{eqnarray} 
where $s_C=s_C^{(A)} \sqcup s_C^{B}$ and $s_{ABC}=s_{ABC}^{(A)} \sqcup s_{ABC}^{B}.$ Then,

\begin{eqnarray}
\partial(r_C \cup R^{(A)}) &=&(A \cup C) \cup(\Gamma \cup s_C^{(B)} \cup s_{ABC}^{(A)}) \label{eq:ssap1} \\
\partial(r_C \cup R^{(B)}) &=&(A \cup B) \cup(\Gamma \cup s_C^{(A)} \cup s_{ABC}^{(B)}) \label{eq:ssap2}
\end{eqnarray}

Then, by RT, Eq.~(\ref{eq:ssap1}) implies that the area of $(\Gamma \cup s_C^{(B)} \cup s_{ABC}^(A))$ is greater than or equal to $S_{AC},$ and Eq.~(\ref{eq:ssap1}) implies that the area of $(\Gamma \cup s_C^{(A)} \cup s_{ABC}^(B))$ is greater than or equal to $S_{BC}.$ Adding these two inequality, and applying RT again, we get the desired inequality

\begin{equation}
2 E_W[AC:BC] \geq S_{AC}+S_{BC}-S_{ABC}-S_C = I(A:B|C) \geq 0,
\end{equation}
where we used the positivity of the conditional mutual information.

\subsection{Upper bounding holographic tripartite information}
We can also use $E^G_W$ to upper bound holographic tripartite information:
\begin{equation}
I(A:B:C) \leq E^G_W(AC:BC)+E^G_W(AB:BC)+E^G_W(CB:AC). \label{eq:triub}
\end{equation}
We could have pursued an inclusion-exclusion style proof for this, but amusingly one does not have to; this follows from Eq.~(\ref{eq:flipssa}).  Adding three instances of Eq.~(\ref{eq:flipssa}), we get:

\begin{gather}
E^G_W(AC:BC)+E^G_W(AB:CB)+E^G_W(BA:CA) \geq \frac{1}{2} \left( I(A:B|C)+I(A:C|B)+I(B:C|A)\right) \nonumber\\
= S_{AB}+S_{AC}+S_{BC}-\frac{3}{2}S_{ABC}-\frac{1}{2}\left(S_A +S_B+S_C \right) \nonumber\\
\geq S_{AB}+S_{AC}+S_{BC}-S_{ABC}-S_A -S_B-S_C \nonumber \\
\end{gather}
where in the last line, we used three party subadditivity ($S_{ABC} \leq S_A+S_B+S_C$). We recognize the last expression above as $I(A:B:C)$, thus completing the proof.

\subsection{Upper bounding holographic cyclic information}

Similarly, the following upper bound on $C_k$ can also be derived:
\begin{equation}
\sum_{i=1}^nE^G_W(A_i, A_{i+i}, \dots, A_{i+k}:A_{i+k}, A_{i+k+1},\dots A_{i+n-1}) \geq C_k (A_1, \dots, A_n), \label{eq:flipcyc}
\end{equation}
where, as before, the indices are to be interpreted mod $n,$ and $n=2k+1.$

To prove this, we use Eq.~(\ref{eq:flipssa}) to get:

\begin{eqnarray}
\sum_{i=1}^n E^G_W(A_i, \dots, A_{i+k}:A_{i+k}, \dots A_{i+n-1}) \nonumber\\
\geq \sum_{\text{cyc}} S(A_1 \cdots A_{k+1})-\frac{n}{2} S(A_1 A_2 \dots A_n)-\frac{1}{2} \sum_{j=1}^n S(A_j)  \nonumber\\
\geq \sum_{\text{cyc}} S(A_1 \cdots A_{k+1})-S(A_1 A_2 \dots A_n)- \frac{n-2}{2} S(A_1 A_2 \dots A_n)-\frac{1}{2}\sum_{j=1}^n S(A_j) \nonumber\\
\geq C_{k}(A_1 \dots A_n), \label{eq:fcycder}
\end{eqnarray}
where we have used subadditivity and that 
\begin{equation}
2 \sum_{\text{cyclic}} S(A_1 \dots A_{k}) \geq (n-2) S(A_1 \dots A_n) + \sum_j S(A_j), \label{eq:fcycd1}
\end{equation}
which follows from repeated application of SSA, as we now show. First, pairwise application of SSA to terms of the form $S(A_i, \dots, A_k)$ and $S(A_{k}, \dots, A_{2k-1})$ on the left-hand side gives:
\begin{equation}
2 \sum_{\text{cyclic}} S(A_1 \dots A_{k}) \geq \sum_j S(A_j) + \sum_{\text{cyc}}  S(A_1 \dots A_{2k-1}) \label{eq:fcycd2}
\end{equation}
Now, let $F$ be a purification of $A_1 \dots A_n,$ so that we have $\sum_{\text{cyc}}  S(A_1 \dots A_{2k-1})= \sum_{i} S(A_i A_{i+1}F).$ Applying SSA now to $S(A_{2i}A_{2i+1}F)$ for $i=1,\dots,k$, and to $S(A_{2i-1}A_{2i}F)$ for $i=1,\dots k$ we get
\begin{eqnarray}
\sum_{i} S(A_i A_{i+1}F) \geq (n-1) S(F) + S(A_1 A_n F) + S(A_1 \dots A_{n-1}F)+S(A_2 \dots A_n F) \nonumber\\
\geq (n-2) S(F)=(n-2) S(A_1 \dots A_n). \label{eq:fcycd3}
\end{eqnarray}
Finally, pairwise application of SSA to $S(A_i \dots A_{i+k+1})$ and $S(A_{i+k+1} \dots A_{i+n-1})$ for $i=1$ to $k$ gives
\begin{equation}
\sum_{\text{cyc}}  S(A_1 \dots A_{k+1})  \geq k S(A_1 \dots A_n) + \sum_{i=1}^k S_i +S(A_{k+1} A_{k+2} \dots A_n) \geq (k+1) S(A_1 \dots A_n). \label{eq:fcycd4}
\end{equation}
Combining Eqs.~(\ref{eq:fcycd2}), (\ref{eq:fcycd3}), and (\ref{eq:fcycd4}) yields Eq.~(\ref{eq:fcycd1}).

\subsection{Cyclic $E_W$ Inequalities}

Here we use as a starting point the cyclic entropy inequalities, Eq.~ (\ref{eq:cyc}), to derive cyclic $E_W$ inequalities. Interestingly, as we show in section \ref{sec:Ep}, the inequalities we arrive are not obviously violated for generic quantum states when $E_W$ is replaced by $E_p.$

We first rewrite the cyclic entropy cone inequalities, Eq.~(\ref{eq:cyc}), in terms of  only mutual information as

\begin{equation}
\sum_{i=2}^n I(A_1 A_2 \dots A_{i-1}:A_i) \geq \sum_{\text{cyclic}} I(A_1:A_2 \dots A_{1+k}). \label{eq:Icyc}
\end{equation}
Then, by the upper bound in Eq.~(\ref{eq:Eulb}), the left-hand side of the inequality above can be upper bounded by a combination of $E_W$'s, which gives

\begin{equation}
\sum_{i=2}^n E_W(A_1 A_2 \dots A_{i-1}:A_i) \geq \frac{1}{2} \sum_{\text{cyclic}} I(A_1:A_2 \dots A_{1+k}). \label{eq:Ewcyc}
\end{equation}

\section{Bounding entanglement entropy with $E_p$ \label{sec:Ep}}
\subsection{Generalized $E_p$}
\begin{figure}[h]
    \centering
    \includegraphics[trim={22cm 11cm 0cm 5cm},clip,width=0.55\textwidth]{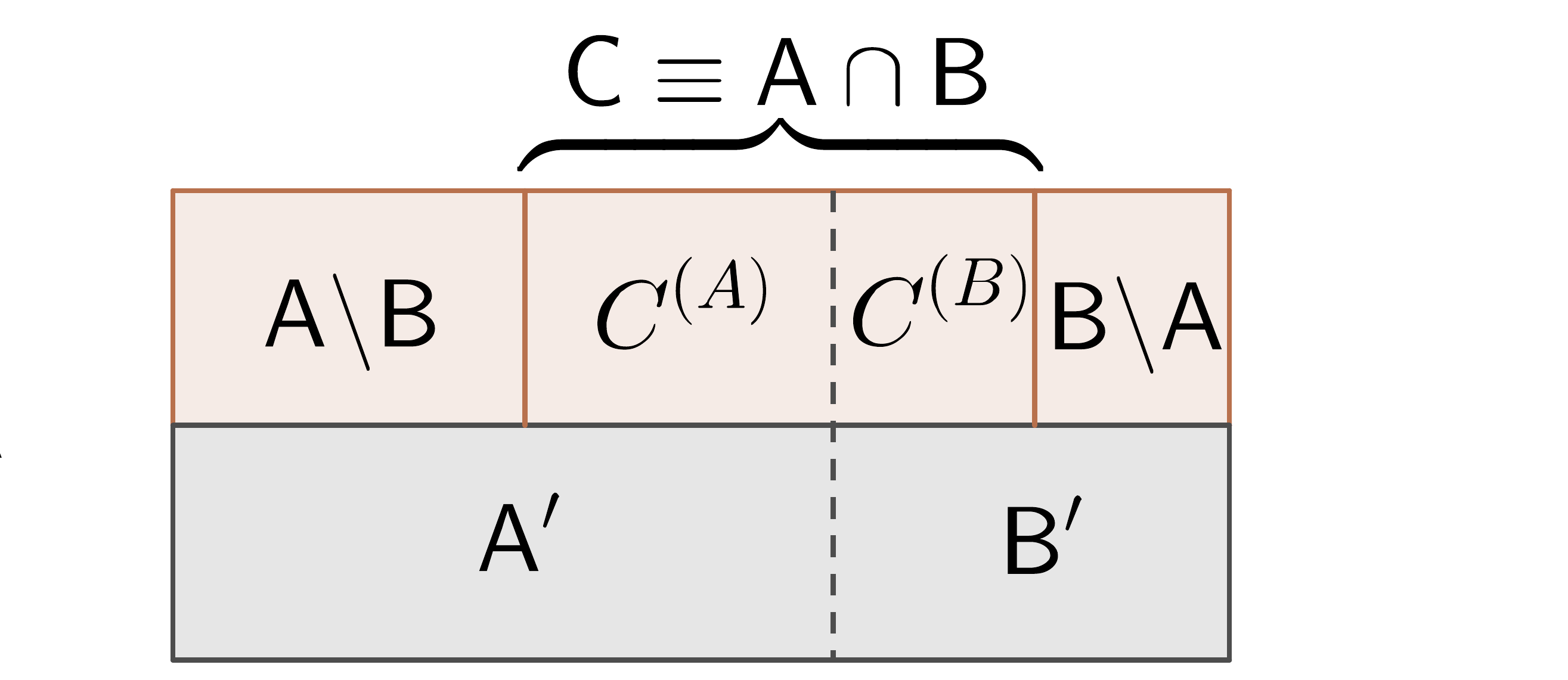}
    \caption{As depicted, for the optimal choices of $A', B'$ and $C^{(A)},$ we have $E^G_p(A:B)=S((A\backslash B)A' C^{(A)}).$}
    \label{fig:EpG}
\end{figure}

Just as we did for $E_W,$ we will similarly need to generalize $E_p$. The {\bf generalized entanglement of purification} $E^G_p$ is defined as
\begin{equation}
E^G_{p}(A:B) = \min_{A\rq{} B\rq{} C^{(A)}} S((A \backslash B) A\rq{} C^{(A)}), \label{eq:EpGdef}
\end{equation}
where as before we require that $AA\rq{}BB\rq{}$ is pure, and we now also require that $C^{(A)} \subset C \equiv A\cap B.$ For convenience, we also define $C^{(B)}=C \backslash C^{(A)}$ (see Fig. \ref{fig:EpG}). Note that the minimization could also have been done over $S((B \backslash A) B\rq{} C^{(B)})$ and also that if $A \cap B = \emptyset$ then $E^G_p(A:B)=E_p(A:B).$

\subsection{$E^G_p$ obeys known $E_p$ inequalities}
We now show that $E^G_p$ also obeys the known inequalities for $E_p$.

The upper bound in Eq.~(\ref{eq:Eulb}) follows from

\begin{eqnarray}
E^G_p(A:B) \leq E_p(A:B \backslash A) \leq \min (S(A),S(B \backslash A) )  \text{ and} \nonumber\\ 
E^G_p(A:B) \leq E_p(A \backslash B :B) \leq \min (S(A \backslash B),S(B)),
\end{eqnarray}
where the first inequality in each line follows from the fact the minimization procedure defining $E^G_p(A:B)$ is less constrained than the one defining  $E_p(A:B \backslash A)$ or $E_p(A \backslash B :B)$\footnote{This is because for  $E^G_p(A:B)$ we are free to choose $C^{(A)},$ while in $E_p(A:B \backslash A)$ we have that $C^{(A)}=\emptyset$, and in $E_p(A \backslash B :B)$ we have $C^{(A)}=C=A \cap B.$}, and the second inequality in each line follows from Eq.~(\ref{eq:Eulb}). Together these imply
\begin{equation}
E^G_p(A:B) \leq \min (S(A),S(B)).
\end{equation}

The lower bound in Eq.~(\ref{eq:Eulb}) follows from 
\begin{eqnarray}
E_p^G(A:B) \geq E_p (A \backslash B:B \backslash A) \nonumber\\
\geq \frac{1}{2} \left( S(A \backslash B)+  S(B \backslash A) - S(A\backslash B \cup B \backslash A) \right) = I^G(A:B),
\end{eqnarray}
where we have used the fact that the minimization procedure for $E_p^G(A:B)$ is more constrained that the one for $E_p (A \backslash B:B \backslash A)$\footnote{This follows from the fact that we can always take $C^{(A)}$ to be part of $A'$ and $C^{(B)}$ part of $B\rq{}.$} and Eq.~(\ref{eq:Eulb}).

For any $C \cap A = \emptyset,$ it is easy to see that 
\begin{eqnarray}
E_p^G(A:BC) \geq E_p (A:B), \text{ and}\\
E_p^G(A:BC) \geq E_P(A-B,(B-A)C) \geq \frac{1}{2} \left( I^G(A:B) + I(A\backslash B:C)\right)
\end{eqnarray}
since adding $C$ further constrains the optimization.

\subsection{Upper bounding conditional mutual information}

We now prove the following upper on conditional mutual information:
\begin{equation}
I(A:B|C) \leq 2 E^G_p(AC:BC). \label{eq:flipssaep}
\end{equation}
Note that, similarly to the $E^G_W$ bound, in the case where $C=\emptyset,$ this reduces to $I(A:B) \leq 2 E_p(A:B)$.

Assume $A\cap B =\emptyset,$ and let $A\rq{}, B\rq,$ and $C^{(A)}\subset C$ define an optimal purification of $(AC:BC)$ according to Eq.~(\ref{eq:EpGdef}). Then, we can get the desired upper bound by repeated application of strong subadditivity:

\begin{gather}
2 E_p^G (AC:BC)+ S(ABC) + S(C) = S(A A\rq C^{(A)}) + S(B B\rq{} C^{(B)})+  S(ABC) + S(C) \nonumber \\
\geq S(A A\rq{} B C) + S(A C^{(A)})+ S(B B\rq{} C^{(B)})+S(C^{(A)}C^{(B)}) \nonumber \\
\geq  S(A A\rq{} B C) + S(B B\rq{} C^{(B)}) + S(C^{(A)}) + S(AC) \nonumber \\
\geq S(BC^{(B)})+S(AC)+S(C^{(A)}) \geq S(BC)+S(AC).
\end{gather}

\subsection{Upper bounding tripartite information and cyclic information}
It is worth noting that the proofs in the previous section of upper bounds for holographic tripartite information, Eq.~(\ref{eq:triub}), and holographic cyclic information, Eq.~(\ref{eq:flipcyc}), depended only on Eq.~(\ref{eq:flipssa}). Since the analogous statement obtained by replacing $E_W^G$ by $E_p^G$, i.e.,  Eq.~(\ref{eq:flipssaep}), also holds, the $E^G_p$ versions of these upper bounds are also true.

\subsection{Lower bound on tripartite information}
One can also extract a quantum lower bound for the tripartite information:
\begin{equation}
I(A:B:C) \geq -2E_p(A:BC)-2E_p(B:C). \label{eq:negmmi}
\end{equation}
This inequality is obviated in the holographic context by positivity of holographic tripartite information. For a general quantum state, however, it is nontrivial. To prove this inequality, we add three instances of positivity of conditional mutual information to find:
\begin{gather}
    I(A:B|C)+I(A:C|B)+I(B:C|A) \nonumber \\
    =2S(AB)+2S(BC)+2S(AC)-S(A)-S(B)-S(C)-3S(ABC)\geq 0.\label{eq:sum3I3}
\end{gather}
We can add to this inequality the inequality
\begin{equation}
S(ABC)-S(A)-S(B)-S(C)\geq -2E_p(A:BC)-2E_p(B:C), \label{eq:triparlbder}
\end{equation}
which follows from two applications of Eq.~(\ref{eq:Eulb}). The sum of Eqs.~(\ref{eq:sum3I3}) and (\ref{eq:triparlbder}) proves the lower bound in Eq.~(\ref{eq:negmmi}).

\subsection{Cyclic $E_p$ inequalities}

If the $E_W=E_p$ conjecture is correct, then for holographic states, it follows from Eq.~(\ref{eq:Ewcyc}) that

\begin{equation}
\sum_{i=2}^n E_p(A_1 A_2 \dots A_{i-1}:A_i) \geq \frac{1}{2} \sum_{\text{cyclic}} I(A_1:A_2 \dots A_{1+k}) \label{eq:Epcyc}
\end{equation}

However, it is interesting to note that in deriving this, we have combined several inequalities, thereby weakening them. For instance, the GHZ state defined by $\ket{\text{GHZ}}=\frac{1}{\sqrt{2}} \left(\ket{0}^{\otimes n} + \ket{1}^{\otimes n} \right)$ is not holographic for $n\geq 4$ and violates instances of Eq.~(\ref{eq:cyc}), but still satisfies Eq.~(\ref{eq:Ewcyc}). This can be seen from the fact that for any $A$ and $B$ disjoint proper subsystems of GHZ, we have\cite{Bagchi}:

\begin{equation}
 E_p(A:B)=S(A)=S(B), \text{ and } I(A:B)=S(A)=S(B).
 \end{equation}
Thus all the terms in Eq~(\ref{eq:Icyc}) and Eq.~(\ref{eq:Epcyc}) are the same, and we can see the former is violated, while the latter satisfied.

Thus, it is plausible that the inequalities in Eq.~(\ref{eq:Epcyc}) hold for general quantum states. Because random states are known to obey the holographic inequalities \cite{Rangamani:2015qwa}, it is also clear that those states would also obey Eq.~(\ref{eq:Epcyc}). 

We now present evidence that these inequalities are also obeyed by W states, which are also known not to be holographic and are defined by 

\begin{equation}
\ket{W}= \frac{1}{\sqrt{n}} \left(\ket{100 \dots 0}+\ket{010 \dots 0}+ \cdots \ket{00\dots 01} \right).
\end{equation}

For any qubit system invariant under the permutation of qubits, we have

\begin{equation}
    \sum_{\text{cyclic}} I(A_1:A_2 \dots A_{1+k}) = n (S(\rho_1)+S(\rho_k)-S(\rho_{k+1})),
\end{equation}
where $\rho_i$ is the reduced density matrix for the i-qubit subsystem. Moreover, by using Eq.~(\ref{eq:Emon}) and permutation symmetry, we can lower bound the left-hand side of Eq.~(\ref{eq:Epcyc}) as follows:

\begin{equation}
    \sum_{i=2}^n E_p(A_1 A_2 \dots A_{i-1}:A_i) \geq \frac{1}{2} \left( 6k S(\rho_1) - 2k S(\rho_2) -S(\rho_{2k}) \right)
\end{equation}.

Thus, Eq.~(\ref{eq:Epcyc}) is implied by 

\begin{equation}
    D \equiv (4k-1) S(\rho_1)-2k(S(\rho_2))-S(\rho_{2k})-(2k+1) S(\rho_k)(2k+1)+S(\rho_{k+1}) \geq 0. \label{eq:Wcheck}
\end{equation}

\begin{figure}[h]
    \centering
    \includegraphics[scale=0.8]{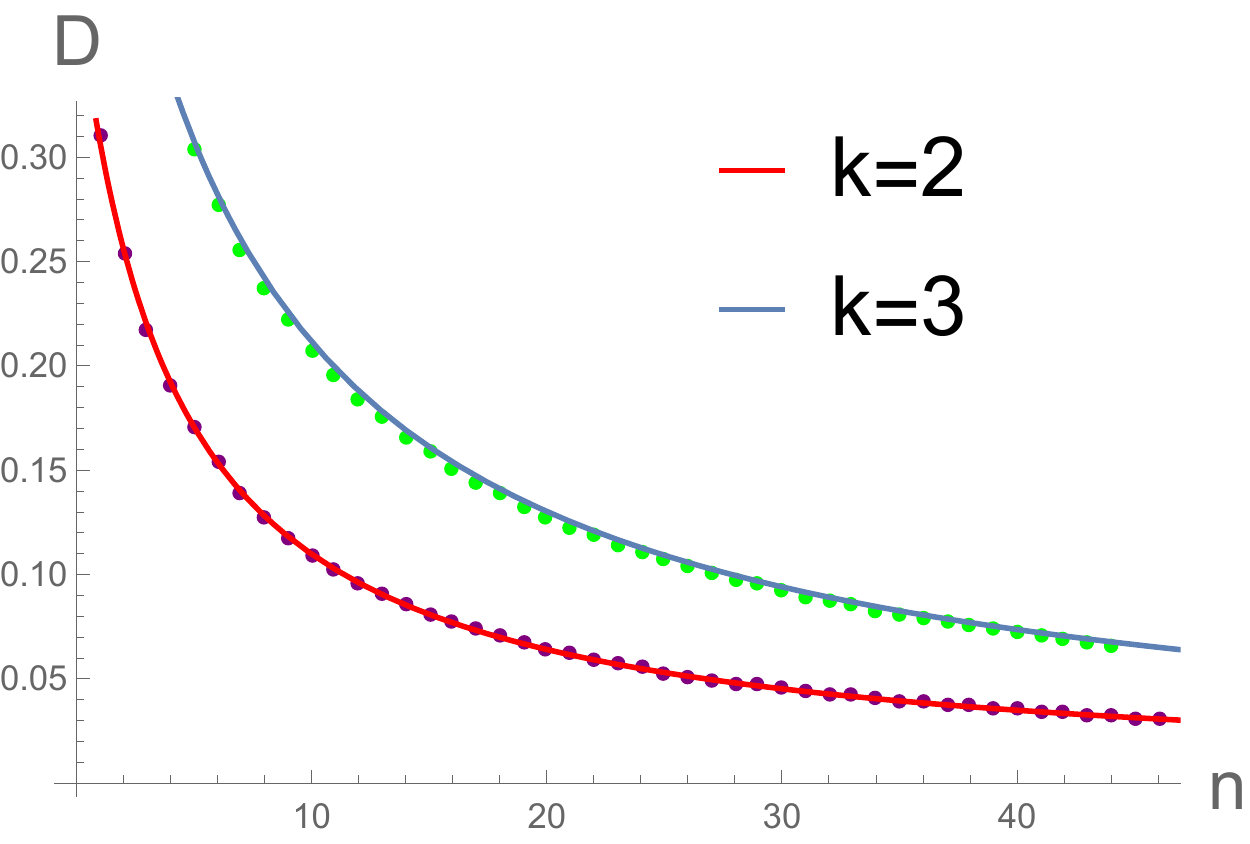}
    \caption{Displaying the left-hand side of Eq.~(\ref{eq:Wcheck}) for $k=2$ and $k=3$ for $W_n$ as a function of $n,$ as well as the best fit curves of the form $D=\frac{B}{n}.$ For $k=2,$ we found $B \approx 1.537,$ and for $k=3,$ we found $B \approx 3.385.$}
    \label{fig:Wplot}
\end{figure}

Let $W_n$ be the density matrix for the  $n$ qubit W state, and let $W_{n,k}$ be its reduced density matrix to a $k$ qubit subsystem. We can now write these in component form as
\begin{equation}
W_n=\frac{1}{n}\left(
\begin{array}{cccccccc}
 0 & 0 & 0 & 0 & 0 & 0 & \cdots & 0 \\
 0 & 1 & 1 & 0 & 1 & 0 & 0 & 0 \\
 0 & 1 & 1 & 0 & 1 & 0 & 0 & 0 \\
 0 & 0 & 0 & 0 & 0 & 0 & 0 & 0 \\
 0 & 1 & 1 & 0 & 1 & 0 & 0 & 0 \\
 0 & 0 & 0 & 0 & 0 & 0 & 0 & 0 \\
 \vdots & 0 & 0 & 0 & 0 & 0 & \ddots & 0 \\
 0 & 0 & 0 & 0 & 0 & 0 & 0 & 0 \\
\end{array}
\right),
\end{equation}
where row $i$ column $j$ contains a $1$ if and only if $i-1$ and $j-1$ are powers of $2,$ and 

\begin{equation}
W_{n,k}=\frac{1}{n}\left(
\begin{array}{cccccccc}
 n-k & 0 & 0 & 0 & 0 & 0 & \cdots & 0 \\
 0 & 1 & 1 & 0 & 1 & 0 & 0 & 0 \\
 0 & 1 & 1 & 0 & 1 & 0 & 0 & 0 \\
 0 & 0 & 0 & 0 & 0 & 0 & 0 & 0 \\
 0 & 1 & 1 & 0 & 1 & 0 & 0 & 0 \\
 0 & 0 & 0 & 0 & 0 & 0 & 0 & 0 \\
 \vdots & 0 & 0 & 0 & 0 & 0 & \ddots & 0 \\
 0 & 0 & 0 & 0 & 0 & 0 & 0 & 0 \\
\end{array}
\right),
\end{equation}
where, apart from the first entry, the same pattern is followed. This allows us to evaluate the left-hand side of Eq.~(\ref{eq:Wcheck}) and verify its positivity (for numerically tractible $n$ and $k$). Moreover, the best fit we found for these curves indicate that this is satisfied for any value of $k$ and $n$ (See Figure \ref{fig:Wplot}).

\section{Future direction: new dictionary entries}

In this section, we use intuition from bit threads \cite{Headrick:2014cta} and from the fact that for states with a holographic dual, $E_p=E_p^\infty=E_{LOq}$ \cite{TakUme} to strengthen the $E_W=E_p$ conjecture. For a definition of these quantities see \cite{Terhal}. For our purposes, we will mainly use the fact that $E_{LOq}(\rho_{AB})$ is roughly equal to the number of EPR pairs needed to get to $\rho_{AB}$ by means of only local operations.

Let $A$ and $B$ be boundary regions, and let the state on it be given by the density matrix $\rho_{AB}.$ Consider the maximum flow $\Phi$ from $A$ and into $B$ \footnote{We can do this by simultaneously maximizing the flow out of both $A$ and $B^c$, which contains $A$, something permitted by the nesting property of bit threads.}. We interpret these as bit threads connecting EPR pairs living on the boundary. The flow lines that leave $A$ and end on $B$ correspond to EPR pairs between $A$ and $B.$ If a thread does not end on $B,$ then it corresponds to an EPR pair between $A$ and the purifying system $A'B'.$ Likewise, the flow lines ending on $B$ and not coming from $A$ are EPR pairs between $B$ and $A'B'.$ We assume that the global boundary state is pure\footnote{This can always be achieved via multiboundary wormhole completion.}.

Given these EPR pairs, the original $\rho_{AB}$ can be reached by local operations on $A$ and on $B.$ Thus, the entanglement of purification is given by the number of bit threads connecting $A$ to $B,$ which is equal to the area of the entanglement wedge cross-section. Note that all lines from $A$ that do not go into $B$ in the flow $\Phi$ do not cross this cross-section (and likewise for lines going into $B$). These facts then motivate the following conjecture. \newline

{\bf Conjecture} Any (non-minimal) surface $\Gamma$ that partitions $r_{AB}$ into a region homologous to $A$ and a region homologous to $B$ is dual to a (suboptimal) purification $A' B'$ such that qubits on the boundary are in $A'$ if they are connected to $A$ without crossing $\Gamma$ or are connected to $B$ while crossing $\Gamma.$ Likewise, boundary qubits are in $B'$ if they are connected to $B$ without crossing $\Gamma,$ or connected to $A$ while crossing $\Gamma.$  \newline

Note that the purifying system of $AB$ dual to $\Gamma$ would be $A'B'$ and $S(AA')=\Phi(\Gamma),$ the flow $\Phi$ across $\Gamma.$ It is clear then that $S(AA')$ is minimized when $\Gamma$ is the entanglement wedge cross-section.

 Combining our conjecture with max cut min flow implies the $E_W=E_p$ conjecture\footnote{In turn, the $E_W=E_p$ conjecture implies Ryu-Takayanagi as a special case.}, as the area of the minimal cross-section of the entanglement wedge is given by the number of bit threads crossing it in this construction, due to its nature as a bottleneck for the flow from $A$ to $B$. Thus, it is clear that for any cut $\Gamma$ there exist an $A'$ and $B'$ such that $\Phi(\Gamma)=S(AA').$ 
 
 Still, the conjecture places nontrivial constraints on the dimensionality of the purifying system. This is because the sum of the number of bit threads emerging from the $A$ system when maximizing the flow through $A$ and those emerging from the $B$ system when maximizing the flow through $B$ upper bounds the log of the dimensionality of the $A'B'$ system. Thus, we get $\log d_{A'B'} \leq S(A)+S(B)$. Note that this is much tighter than the upper bound given in \cite{Terhal}; this is not surprising, however, given that holographic states have much less entanglement than the generic quantum states considered in \cite{Terhal}. Moreover the overall holographic state is pure, and thus one does not get confounding entanglement of purification from considering classical mixtures.

It would be interesting to study the plausibility of this conjecture in toy models of holography, in particular suitable generalizations of the qutrit code \cite{Almheiri:2014lwa}, perfect tensors \cite{Pastawski:2015qua}, or the random tensor model \cite{Hayden:2016cfa}. In such simpler models, it may be possible to explicitly reconstruct $A'$ and $B'$ from given known $A$ and $B$.

Other interesting directions of future research include to either prove as disprove Eq.~(\ref{eq:Epcyc}) as an inequality valid for all quantum systems, and to extend the results of the present paper to the fully covariant case.

\section{Conclusion}

In this paper, we have considered upper and lower bounds for several information theoretic quantities, including bounds on the conditional mutual information, tripartite information, and cyclic information. Despite being motivated by holography, we have shown these to hold for all quantum states. We have also found a new family of holographic inequalities for $E_W,$ and provided evidence that the corresponding inequality for $E_p$ may be true for all quantum states. These results are summarized in Table \ref{tab:res}.

Finally, we conjectured a potential extension of the $E_W=E_p$ conjecture of \cite{TakUme, Nguyen}, which asserts that all cuts of the entanglement wedge are dual to purifications. If true, it may be possible to write such a map explicitly, which could lead to an efficient way of computing entanglement of purification.

\begin{table}[h]
\label{framestable}
\begin{center}
\scalebox{0.8}{
\begin{tabular}{|c|c|c|}
\hline
 & Lower Bound   &  Upper Bound\\ \hline
Mutual Information & ${\mathbf 0}$ & $\mathbf{2E_p(A:B)}$  \\
I(A:B)&  &  \\ \hline
Conditional Mutual Information & $\mathbf 0$  & $\mathbfit{2E^G_p (AC:BC)}$ \\
I(A:B|C)&  & \\ \hline
Tripartite Information & $0, $ and & $\mathbfit{E^G_p(AC:BC)+E^G_p(AB:BC)}$ \\ 
I(A:B:C)& $\mathbfit{-2E_p(AB:C)-2E_p(B:C)}$& $\mathbfit{+E^G_p(AB:AC)}$\\ \hline
Cyclic Information&  $0$ &\footnotesize{$\mathbfit{\sum_{i=1}^n E^G_p(A_i, \dots, A_{i+k}:A_{i+k},\dots A_{i+n-1})}$} \\ 
$C_k(A_1, \dots, A_{2k+1})$ & & \\ \hline
\end{tabular}}
\caption{The main results are listed in the table. Results that, to the best of the authors knowledge, are new are in \emph{italics} (the upper bound on conditional mutual information was proved for $E_W$ in the language of bit threads in \cite{Headrick:2014cta}). In {\bf bold} are the inequalities known to hold for all quantum states, while the others hold holographically, but are known to be violated in general. In addition to these results, we have also shown the holographic inequality in Eq.~(\ref{eq:Ewcyc}), which does not fit neatly into  this table. The general version of this inequality, pending the $E_W=E_p$ conjecture, is given by Eq.~(\ref{eq:Epcyc}). We have shown this to hold for several non-holographic states, but its general validity is still an open question.} \label{tab:res}
\end{center}
\end{table}

\section*{Acknowledgements}
We would like to thank Raphael Bousso, Veronika Hubeny, Adam Levine, Rob Myers, Daniel Parker, and Sean Weinberg for discussions. We thank Linghang Kong for comments and pointing out typos in an earlier version of this work. This work is supported in part by the Berkeley Center for Theoretical Physics. N.B. is supported by the National Science Foundation, under grant number 82248-13067-44-PHPXH, and I.F.H. is supported by National Science Foundation under grant PHY-1521446.

\bibliographystyle{utcaps}
\bibliography{all}
\end{document}